\documentclass[12pt]{article}
\usepackage{xspace,amssymb,amsmath,helvet,graphicx}

\newcommand{\eg}{e.g.\xspace}
\newcommand{\ie}{i.e.\xspace}

\newcommand{\etc}{etc.\@\xspace}

\newcommand{\role}{r\^{o}le}

{\begin{list}{}{%
\settowidth{\labelwidth}{\textsf{#1:}}%
\setlength{\leftmargin}{\labelwidth}
\addtolength{\leftmargin}{\labelsep}
\setlength{\itemindent}{0pt}
}}%
{\end{list}}
\begin{document}
\title{Discrete distributions from a Markov chain}
\author{Rose Baker\\School of Business\\University of Salford\\The Crescent\\Salford UK}
\maketitle

\begin{abstract}A discrete-time stochastic process derived from a model of basketball is used to generalize any discrete distribution.
The generalized distributions can have one or two more parameters than the parent distribution. Those derived from binomial, Poisson 
and negative binomial distributions can be underdispersed or overdispersed. 
The mean can be simply expressed in terms of model parameters, thus making inference for the mean straightforward.
Probabilities can be quickly computed, enabling likelihood-based inference. Random number generation is also straightforward.
The properties of some of the new distributions are described and their use is illustrated with examples.

\end{abstract}
\section*{Keywords}
Binomial distribution; Poisson distribution; underdispersion; overdispersion; stochastic process; basketball
\section{Introduction}
Discrete distributions are used when modelling count data, and in particular the dependence of counts on covariates.
There is a wide range of discrete distributions and of application areas, \eg  economics, life sciences, reliability (\eg Johnson {\em et al}, 2005).

The Poisson and binomial distributions are the best-known discrete distributions. 
However, count data often show overdispersion or more rarely underdispersion relative to these, and the probability of occurrence of zero events 
often differs from what these distributions would predict. Hence many generalizations of these distributions have been developed.

The stochastic model described here gives rise to a discrete distribution defined on the non-negative integers that can generalize any discrete distribution
and allows underdispersion and overdispersion. 
Probabilities can be quickly computed, an important property for practical use.
Because its mean has a simple form, inference that requires modelling the mean as a function of covariates
is straightforward; for example, the average marginal effect of changing a covariate value is immediately known from the regression coefficient.
This is an attractive property. Without it, often a scale parameter can be regressed on covariates (\eg Kharrat {\em et al} 2019), but the interpretation of the results is then more complex.

The stochastic process occurs in discrete time, and is a sequence of 
Bernoulli trials where the probability of a success is $r_1$ following a failure,
and $1-r_2$ following a success, \ie the process is a Markov chain. Initially, it is in the stationary state. The process is stopped at some fixed or random time $N$.

A referee pointed out to the author that this distribution already exists: it is the Markov binomial distribution (\eg Dekking and Kong, 2011).
However, this article presents some new material, and so is arguably worth reading.

This process and the resulting discrete distribution was explored as a model of the number of goals scored by a team in the game of basketball
when $N$ goals have been scored in total.
Often, new statistical tools are developed in the process of analysing data. This happened here, where a new distribution was developed 
in the course of ongoing research into netball and basketball. It was realized that the distribution of team 1 goals
had much wider applicability, and hence this article is a spin-off from that applied research.

In basketball, after a team scores a goal (a basket, 2 or 3 points) the ball is turned over to the opposition.
Under a simple model, let team 1 and team 2 have probabilities $r_1, r_2$ respectively of scoring when the ball has been turned over to them and they have starting possession.
Hence team 1 has probability $1-r_2$ of scoring when team 2 has starting possession.

This rule of basketball gives rise to a stochastic process in discrete time, where the time step is the scoring of a goal by either team, and the events of interest are scoring of goals by team 1.
In a full model of basketball, goals would be 3-point baskets with some probabilities $t_1, t_2$, but this is irrelevant for the purpose here and only goals are considered.
Clearly one would expect that a team could score more easily when it had starting possession, so that  $r_1 > 1-r_2$
or $r_1+r_2 > 1$.
In practice $r_1, r_2 \simeq 0.8$. Here, to obtain a wider class of models, this constraint is relaxed.

The process stops with probability $P_n$ after $N$ goals have been scored in total. This number $N$ can be fixed or follow any discrete distribution
on the non-negative integers (the `parent' distribution). In basketball, by the way, it seems to be roughly Poisson.

The times of scores by team 1 form a renewal process. The probability that team 1 has starting possession has an equilibrium or stationary probability $s$.
If it is $s_{i-1}$ at time $i-1$, at time $i$ it is
\[s_i=(1-r_1)s_{i-1}+r_2(1-s_{i-1}),\]
as team 1 either keeps possession on failing to score, or acquires it when team 2 scores.
Equating $s_{i-1}$ to $s_i$ yields $s=r_2/(r_1+r_2)$. Starting the stochastic process off with the probability $s$ that team 1 has starting possession
means that times of team 1 starting possession form an asynchronous renewal process, and we shall see that the mean number of team 1 goals is therefore 
\begin{equation}\mu=\frac{r_1 \text{E}(N)}{r_1+r_2}\label{eq:mu}\end{equation}

\begin{figure}[!htb]
\centering
\makebox{\includegraphics{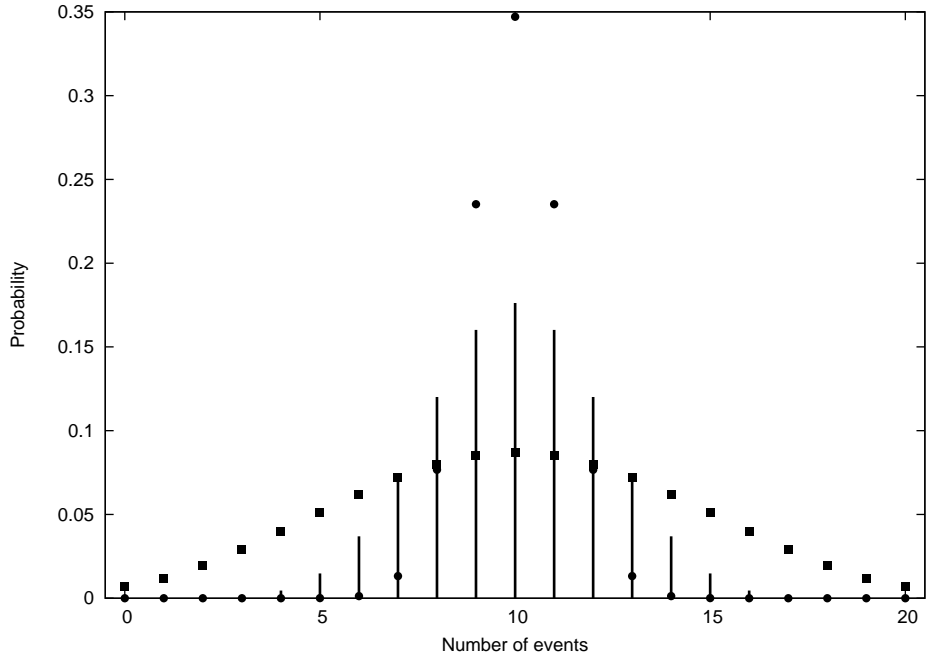}}
\caption{\label{figb}Binomial and over and under-dispersed distributions for $n=20$.
For the binomial, $r_1=r_2=0.5$, for overdispersed $r_1=r_2=0.2$, for underdispersed $r_1=r_2=0.8$.}
\end{figure}

When $r_1=1, r_2=0$, team 1 always scores, and we regain the parent distribution. Hence this stochastic process generalizes any discrete distribution.
When $1-r_2 > r_1$, \ie $r_1+r_2 < 1$, a goal scored by team 1 loses possession and results in a greater probability that team 1 scores at the next time step.
This is a self-exciting process that will have high variance. Conversely, if $1-r_2 < r_1$ or $r_1+r_2 > 1$ team 1 is less likely to score at the next step, and
we have a self-dampening process with lower variance.
This is illustrated in figure \ref{figb}.

Using a Poisson parent distribution, the distribution perhaps most similar to what is proposed here in concept is the `burnt-fingers' distribution, \eg Arbous and Kerrich (1951), Johnson {\em et al} (2005).
In that distribution the first event in a Poisson process occurs with a different intensity to later events. However, in the model used here,
events occurring with a different probability can be triggered multiple times. 

\begin{figure}
\centering
\makebox{\includegraphics{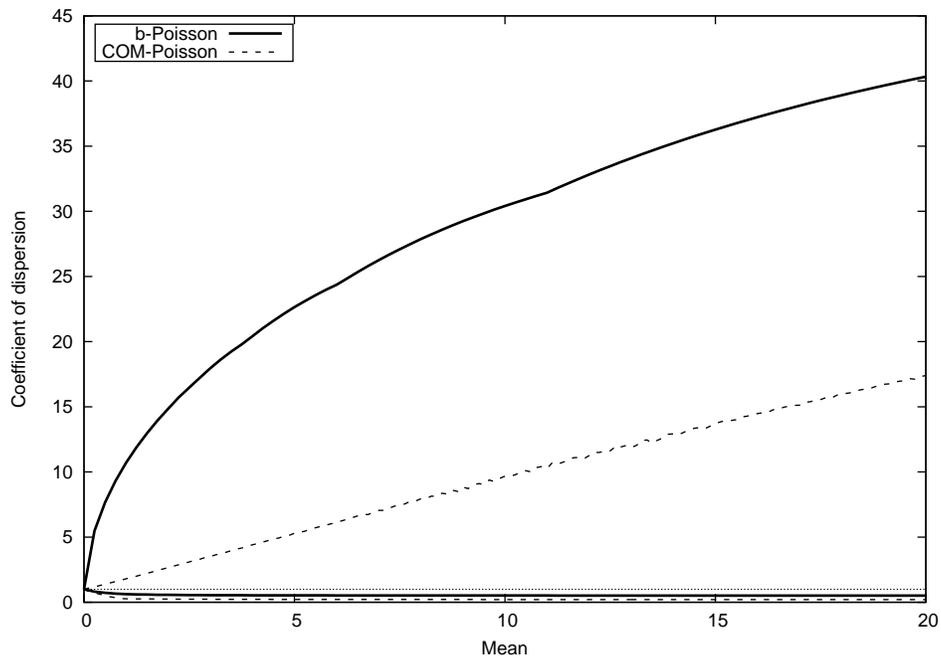}}
\caption{\label{figa}Maximum and minimum coefficients of dispersion $\sigma^2/\mu$ against mean $\mu$ for the b-Poisson and COM-Poisson distributions.}
\end{figure}

A more practical comparison is with the COM-Poisson distribution, which is currently popular for modelling both underdispersed and overdispersed data
(Shmueli {\em et al}, 2005). The probability mass function (pmf) is $P_n \propto (\mu^n/n!)^{\nu}$ where $\nu > 0$. The range of coefficients of dispersion (variance over mean) that can be achieved with 
the Poisson-like distribution and the COM-Poisson distribution is shown in figure \ref{figa}.

The Poisson-like distribution proposed here can achieve much higher variances for a given mean, while the COM-Poisson can achieve lower variances.

The distributions proposed here are however not tied to a Poisson parent, and for example also provide over and underdispersed binomial-like distributions.
An underdispersed generalization of the binomial distribution is useful \eg for modelling the number of embryos
produced from some number of ovulations in sheep \etc
The idea used here of using a stationary renewal process to obtain a simple expression for the mean of a discrete distribution is not new and can be found in Baker (2018).

It is proposed to call the new distributions b-binomial, b-Poisson \etc (b for basketball), and the basketball example is used throughout to fix ideas. The next section discusses the distributions in more detail, starting with the calculation of probabilities.
\section{Simple Calculation of probabilities}
The methods here are useful when it is desired to compute many probabilities. A faster method for computing the single probabilities needed for likelihood-based inference is given later.

Let the distribution of the total number of goals scored (by either team) have pmf $P_n$, for $n$ a non-negative integer.
Let the probability that team 1 has scored $i$ goals and has starting possession when $n$ goals in total are scored be $q_{in}^{(1)}$, and similarly for team 2.
Then
\[q_{in}^{(1)}=(1-r_1)q_{i,n-1}^{(1)}+r_2 q_{i,n-1}^{(2)},\]
\[q_{in}^{(2)}=r_1q_{i-1,n-1}^{(1)}+(1-r_2) q_{i-1,n-1}^{(2)}.\]
Initially, $q_{00}^{(1)}=r_2/(r_1+r_2)$, $q_{00}^{(2)}=r_1/(r_1+r_2)$, and all probabilities for $i > 0$ are zero.
The required probabilities $p_{i}$ are then
\begin{equation}p_{i}=\sum_{n=0}^\infty (q_{in}^{(1)}+q_{in}^{(2)})P_n.\label{eq:comp}\end{equation}

The calculation can be done symbolically on a computer for the case where $N$ is fixed, by keeping track of the powers $a, b, c, d$ in terms of the form $r_1^a(1-r_1)^b r_2^c (1-r_2)^d$.
Expressions can be simplified by combining terms with the same values of $a, b, c, d$, and, when the coefficients are equal,
by setting
\[r_1^{a+1}(1-r_1)^b r_2^c (1-r_2)^d+r_1^a(1-r_1)^{b+1} r_2^c (1-r_2)^d \rightarrow r_1^a(1-r_1)^b r_2^c (1-r_2)^d,\]
and similarly for $r_2$. Results of this computer algebra calculation for small values of $N$ are shown in appendix A.
This recursive computation can also be done numerically. 
For a distribution of $N$, the principle of the computation is to compute the $q_{in}$ and to use (\ref{eq:comp}) to accumulate the $p_i$, until values of $P_n$ are tiny in the tail.
Some code for this is given in the supplementary materials.

It follows from appendix A or probabilistically that for $N$ fixed at $n$, probabilities can be derived analytically and that in general
\begin{equation}p_0=(r_1+r_2)^{-1}r_2(1-r_1)^{n-1}\label{eq:p0}\end{equation}
\begin{equation}p_1=(r_1+r_2)^{-1}\{2r_1r_2(1-r_1)^{n-2}+(n-2)r_1r_2^2(1-r_1)^{n-3}\},\label{eq:p1}\end{equation}
where the formula for $p_0$ applies for $n > 0$ and the formula for $p_1$ applies for $n > 1$; when $n=1$, $p_1=(r_1+r_2)^{-1}r_1$.

For fixed $N$, the symmetry
\begin{equation}p_i(r_1,r_2)=p_{n-i}(r_2,r_1)\label{eq:symm}\end{equation}
follows from considering goals won by one team as goals conceded by the other.
\section{Properties}
\subsection{The mean: relation to renewal processes}
The `times' (numbers of goals from either team) at which team 2 (say) start with possession form a discrete renewal process,
with probabilities $P_j$ for a sequence of length $j$ and distribution function $F_j$ given by 
\begin{align}P_1&=1-r_2,\nonumber \\
P_{m > 1}&=r_1r_2(1-r_1)^{m-2},\nonumber \\
F_{m>1}&=1-r_2(1-r_1)^{m-1}\label{eq:ren}
\end{align}
so that team 2 either fail to score and regain possession immediately,
or score, then team 1 fails to score $m-2$ times, and then scores, returning possession to team 2. This distribution is a mixture of 1 and a shifted geometric distribution, as only $P_1$ differs from 
geometric.
Each renewal counts 1 goal for team 1.
The pgf (probability generating function) is
\begin{equation}G(z)=(1-r_2)z+z^2r_1r_2/(1-(1-r_1)z),\label{eq:g}\end{equation}
because the pgf of the geometric distribution starting at 2 is $r_1z^2/(1-(1-r_1)z)$.

Let $T$ be the time (number of goals by either side) for team 1 to score when team 2 is initially in possession.
From (\ref{eq:g}), the mean time to score for team 1 when team 2 is in possession is $\text{E}(T)\equiv m_1=1+r_2/r_1$ and the variance $\text{var}(T)\equiv v_1=(2-r_1-r_2)r_2/r_1^2$.
The sequences of team 1 and team 2 possession also form an alternating renewal process, so the probability that team 1 is in possession at a random point in time is $m_1/(m_1+m_2)=r_2/(r_1+r_2)$.
This was shown more simply in the introduction.
The third moment $\text{E}(T^3)=1+r_2(r_1^2+6)/r_1^3$. From this the third central moment is $\kappa_1=\text{E}(T^3)-3m_1v_1-m_1^3$.

Since each renewal is a goal for team 1, the number of goals scored by team 1 is the renewal function at time $N$.
With an asynchronous renewal process,  team 1 starts in possession with probability $r_2/(r_1+r_2)$.
The formula for the mean team 1 number of goals is then exactly $\mu=\text{N}/\text{E}(T)=r_1\text{E}(N)/(r_1+r_2)$.

The distribution of time to team 1 scoring from the equilibrium state is
\begin{align}P_1&=\frac{r_1}{r_1+r_2},\nonumber \\
P_{m > 1}&=\frac{r_1r_2}{r_1+r_2}(1-r_1)^{m-2}\nonumber \\
F_m&=1-\frac{r_2}{r_1+r_2}(1-r_1)^{m-1}.\label{eq:ren0}
\end{align}

\subsection{Numbers of parameters and their ranges}
Besides the parameters of the `parent' distribution, we have $r_1, r_2$, adding two parameters.
However, if the parent distribution is binomial, Poisson, or negative binomial, in the case that
$r_1=1-r_2$ (and so goals are scored with constant probability $r_1$), we have no extra parameters, because we then have merely a thinning of the 
distribution of $N$  with mean reduced to $r_1\text{E}(N)$. For the overdispersed b-Poisson distribution with $r_1+r_2 < 1$, as $r_1$ changes, a very similar (but not identical) distribution can be found
on varying $r_2, \text{E}(N)$. Hence for overdispersed distributions, $r_1$ can be set to unity without unduly spoiling the model fit to data.

Parameter ranges of course include $0 < r_1 < 1$, $0 < r_2 < 1$, as $r_1, r_2$ are probabilities.
However, the end points $0, 1$ can be included for one of these probabilities.
\subsection{The abundance of new distributions}
Any distribution can be chosen as the parent distribution. The simplest choice is that $N$ is fixed at $n$, so that $P_n=1$.
Then if $r_2=1-r_1$ the new distribution is $\text{Bin}(r_1, n)$, otherwise it generalizes the binomial with one extra parameter,
and the resulting distribution can be over or under-dispersed. As $N \rightarrow\infty$ and $r_1 \rightarrow 0$ such that $r_1N\rightarrow \mu$,
when $r_2=1-r_1$ we have the Poisson distribution, and otherwise a 1-parameter generalization of the Poisson distribution that must be overdispersed, because $r_1+r_2 < 1$.

\begin{figure}
\centering
\makebox{\includegraphics{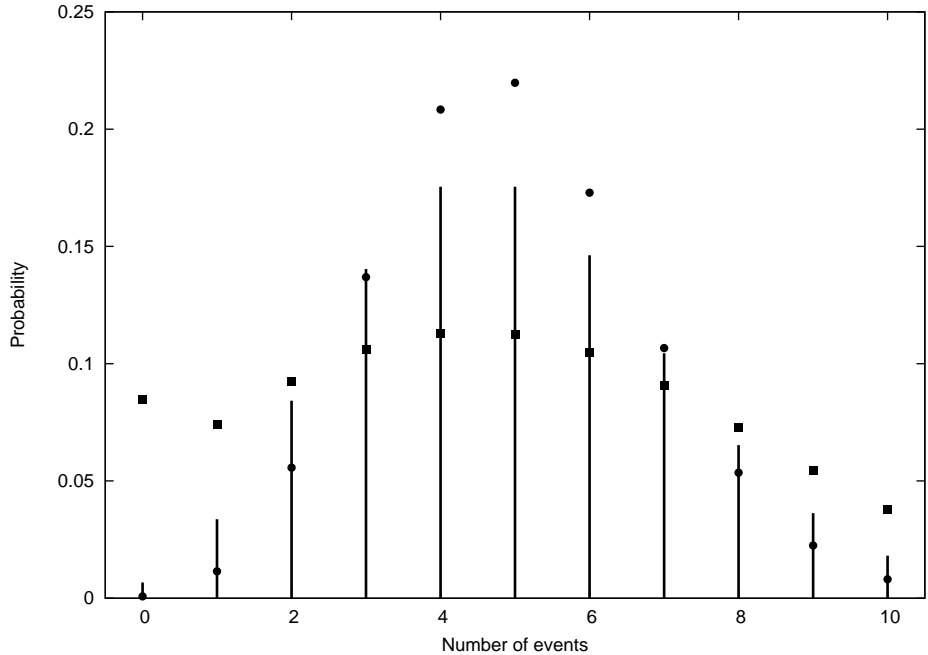}}
\caption{\label{figc}Poisson and over and under-dispersed distributions for $\mu=10$.
For the Poisson, $r_1=r_2=0.5$, for overdispersed $r_1=r_2=0.2$, for underdispersed $r_1=r_2=0.8$.}
\end{figure}

Choosing a fixed value of $N$ as the parent  distribution gives a 3-parameter generalization of the binomial, with one parameter more than the binomial distribution,
that can be over or under-dispersed.
Choosing the Poisson distribution as parent gives a 3-parameter generalization of the Poisson ( 2 extra parameters) that can also be over or under-dispersed
(figure \ref{figc}).
In general, a parent distribution gives rise to a new distribution, and this could be the parent distribution for further distributions.
\subsection{Modes}
No theoretical work has been done on this, but from observation the b-binomial distribution can be unimodal, bimodal or trimodal when overdispersed, with a spike at zero and/or the maximum.
The b-Poisson distribution can also have a spike at zero when overdispersed.
\subsection{Relation to hurdle models}
In a hurdle model, events occur once a threshold or hurdle is cleared, so for example a couple may decide whether or not to have children, but once they have one child, 
the hurdle is cleared and they may well have more.
In the model proposed here, once team 1 has scored a goal, team 2 has possession, and so the probability of further goals being scored may increase or decrease.
However, once team 1 again has possession, this is a regeneration point for the stochastic process. Hence the hurdle is never finally cleared in this model.

\subsection{Calculable probabilities}
For the binomial and Poisson cases, the probabilities $p_0, p_1$ can be derived analytically with some tedious algebra; probabilities have already been given in appendix A
for the degenerate binomial case of $N$ fixed at $n$.
Let the distribution of the total number of goals be $\text{Bin}(r_3,N)$.
Then from (\ref{eq:p0})
\[p_0=\frac{r_2}{(r_1+r_2)(1-r_1)}\{(1-r_1r_3)^N-(1-r_3)^N)\}+(1-r_3)^N,\]
and on taking the limit, for the b-Poisson case where the total number of goals has mean $\mu_0$,
\[p_0=\frac{r_2}{(r_1+r_2)(1-r_1)}\{\exp(-r_1\mu_0)-\exp(-\mu_0)\}+\exp(-\mu_0).\]
Similarly, for the binomial case
\begin{align*}
p_1&=(r_1+r_2)^{-1}\{\frac{2r_1r_2(1-r_1r_3)^N}{(1-r_1)^2}+\frac{r_1r_2^2Nr_3(1-r_1r_3)^{N-1}}{(1-r_1)^2}-\\
&\frac{2r_1r_2^2(1-r_1r_3)^N}{(1-r_3)^3)}+Nr_1r_3(1-r_3)^{N-1}-\frac{2r_1r_2(1-r_3)^N}{(1-r_1)^{2}}\\
&-\frac{2r_1r_2^2(1-r_3)^N}{(1-r_1)^{3}}-\frac{2r_1r_2r_3N(1-r_3)^{N-1}}{1-r_1}+\frac{r_1r_2^2r_3(1-r_3)^{N-1}}{(1-r_1)^{2}}\}.\\
\end{align*}
In the Poisson limit,
\begin{align*}
p_1&=(r_1+r_2)^{-1}\{\frac{2r_1r_2\exp(-r_1\mu_0)}{(1-r_1)^2}+\frac{r_1r_2^2\mu_0\exp(-r_1\mu_0)}{(1-r_1)^2}-\\
&\frac{2r_1r_2^2\exp(-r_1\mu_0)}{(1-r_1)^3}+r_1\mu_0\exp(-\mu_0)-\frac{2r_1r_2\exp(-\mu_0)}{(1-r_1)^2}+\frac{2r_1r_2^2\exp(-\mu_0)}{(1-r_1)^3}-\\
&\frac{2r_1r_2\mu_0\exp(-\mu_0)}{1-r_1}+\frac{r_1r_2^2\mu_0\exp(-\mu_0)}{(1-r_1)^2}\}.\\
\end{align*}
Being able to compute $p_0$ means that inference could be done for modified models where the zero probability is changed, as the mean is still analytically calculable
as a function of model parameters.
If $p_0 \rightarrow p_0^\prime$, probabilities $p_i \rightarrow p_i/(1+p_0^\prime-p_0)$ for $i > 0$ and the mean $\mu \rightarrow \mu/(1+p_0^\prime-p_0)$.

Being able to calculate $p_1$ means that one can test analytically for a spike at zero where $p_0 > p_1$.
\subsection{Random number generation}
Random numbers can be easily generated by simulating who has original possession, and then the total number of goals from the parent (\eg Poisson) distribution, and then
simulating scoring by teams 1 and 2 in a basketball match. This assumes that one can generate random numbers from the parent distribution.
A faster method is to generate times of events (team 1 goals) from distributions (\ref{eq:ren}) and (\ref{eq:ren0}) by using the inverse-F method.
An algorithm is given here, where the total number of goals is $n$:
\begin{enumerate}
\item set $c=r_2/(r_1+r_2)$, set nsum, irand to zero;
\item generate a uniform random number $U$;
\item if $U > c$ set $X=1$;
\item else set $X=\text{floor}\{\ln(U/c)/\ln(1-r_1)\}+2$;
\item increase nsum by $X$ and finish if nsum exceeds $n$;
\item increase irand by 1;
\item repeat from step 2, using $c=r_2$ for this and subsequent generations of $X$.
\end{enumerate}
This gives random numbers from the b-binomial distribution, and it is easy to then generate from \eg the b-Poisson, by generating $N$ from a Poisson distribution.
The code can be optimised from the simple version given here. For example, $\ln(U/c)=\ln(U)-\ln(c)$, and $-\ln(U)$ is exponentially distributed.
A fast method for generating exponential random numbers due to Ahrens and Dieter could be exploited, described with a correction in Gentle (2003).

These relatively computationally simple methods are often what is required.
If generating many random numbers, generation could be speeded up by using the standard methods for discrete distributions (see \eg Gentle, 2003),
where all probabilities are initially computed.
\subsection{Asymptotic probability distribution for large mean}

Noortwijk and van der Weide (2006) show that the variance of the number of renewals $\phi_1^2 \simeq  N\text{var}(T)/\text{E}(T)^3$, and the number of 
events (goals) $N_1$ is asymptotically normal.
Cox (1962) gives a correction for the variance of a (continuous)  asynchronous renewal process: the term to be added to $\phi_1^2$ is $\xi_1=1/6+v_1^2/2\mu_1^4-\kappa_1/3\mu_1^3$.
The total number of goals $N$ is a random variable, and so the variance of the number of renewals tends to $\text{E}(N)\text{var}(T)/\text{E}(T)^3+\text{var}(N)/\text{E}(T))^2+\xi_1$.
Only the mean and variance of $N$ are needed.

Thus the distribution of the number of team 1 goals is asymptotically normal $\text{N}[\mu,\phi_1^2]$ , with the mean known exactly and variance known to a good approximation in terms of model parameters.
This approximation could be useful in calculating probabilities for very large counts or calculating probabilities for censored data, where the count is known only to exceed some large number.
\subsection{Fast computation of single probabilities}
The method of probability computation mentioned earlier is simple but cannot be claimed to be fast. For binomial-type distributions,
producing the $i$th probability from a generalized distribution takes $O(in)$ multiplications, $O(\min(i,n-i)n)$ by exploiting the symmetry (\ref{eq:symm}). For a generalized Poisson or other general discrete distribution, the computation takes longer,
because each parent probability $P_n$ gives rise to $O(n)$ operations, and one must continue until $P_n$ is tiny. Hence if $m$ Poisson probabilities are not tiny, the number of multiplications is $O(im)$.

An algorithm using De Pril's method (De Pril, 1985) and exploiting the scaling of geometric probabilities can however deliver single probabilities in time $O(m-i)$. 
It is briefly described here, with code given in the supplementary materials.
The strategy is to compute the pmf of the time to the $i$th event and then apply the result that the probability of at least $i$ events is the probability that $s$ does not exceed $n$.

De Pril's convolution formula for the $i$th convolution of a discrete distribution with support $\{1,2,3\cdots\}$ and pmf  $P_1(1),P_1(2)\cdots$ is a special case of a formula given in De Pril (1985):
\[P_i(i)=P_1(1)^i,\]
then
\[P_i(s)=P_1(1)^{-1}\sum_{x=1}^{s-i}\{\frac{i+1}{s-i}x-1\}P_1(x+1)P_i(s-x)\]
recursively for $s=i+1, i+2\cdots$.

This is applied for the probabilities in (\ref{eq:ren}) for the $i$th and $i-1$th convolutions,
then we also apply the final ($i$th) convolution using probabilities from (\ref{eq:ren0}),
we write finally for the required probability
\[p_i=\sum_{n=i}^m\sum_{s=i-1}^{n-1}\{P_{i-1}(s)-P_i(s)\}\{1-\frac{r_2}{r_1+r_2}(1-r_1)^{n-s-1}\},\]
where $m$ is the highest probability that is not negligible.

Despite De Pril's wonderful algorithm, computing time would still not be linear in $m$. However, the geometric nature of the probabilities in (\ref{eq:ren}), (\ref{eq:ren0})
allows this linearity.

Let $T_{as}=0, T_{bs}=0$ and update using the formulae
\[T_{as}=(1-r_1)T_{a,s-1}+P_1(2)P_i(s-1),\]
\[T_{bs}=(1-r_1)(T_{b,s-1}+T_{a,s-1})+P_1(2)P_i(s-1).\]
Then De Pril's formula becomes
\[P_i(s)=P_1(1)^{-1}\{\frac{i+1}{s-i}T_{bs}-T_{as}\}.\]

Finally, set $A_{i-1}=B_{i-1}=0$ and update
\[A_n=A_{n-1}+P_{i-1}(n)-P_i(n),\]
\[B_n=(1-r_1)B_{n-1}+P_{i-1}(n)-P_i(n).\]
Then 
\[p_i=\sum_{n=i-1}^m P_n(A_n-\frac{r_2}{r_1+r_2}B_n).\]
Hence as summation starts at $i-1$ and continues to $m$, the number of operations is $O(m-i)$.
A numerical instability was found, which occurs when the required probability is a low one, and the mean of the distribution is high.
Then as the algorithm processes high values of $n$, many tiny terms are added to the probability $p_i$, which eventually starts to oscillate increasingly.
This is easily eliminated by stopping as soon as the next contribution $A_n-\frac{r_2}{r_1+r_2}B_n$ to the probability would be negative.

This computation speed is comparable to that for the popular COM-Poisson distribution (Shmueli {\em et al}, 2005).
In fact, it is several times faster, because although it requires a lot of multiplications and a few divisions, the COM-Poisson computation requires a lot of exponentiations,
which are more expensive.

\section{Examples}
Besides its use in ongoing basketball research, the b-Poisson model was fitted to two datasets.
The first is the completed fertility dataset from the second (1985) wave of the German Socio-Economic Panel, described in Winkelmann (1995). It contains number of children (0-11)
and 10 demographic covariates for 1243 women. The count distribution is slightly underdispersed, and becomes more so after regressing on the covariates.
The second is from Hilbe (2011) who fits the negative binomial distribution to a number of datasets. One is the `affairs' dataset, with 601 observations
from Fair (1978), reporting counts of extramarital affairs over a year in the USA. This distribution is overdispersed, with a big spike at zero containing 75\% of observations.
For both datasets,  the mean of the Poisson distribution of $N$ was taken as $\alpha\exp({\boldsymbol \beta}^T{\bf x})$, where ${\bf x}$ is the vector of covariates, ${\boldsymbol \beta}$ a vector of coefficients,
and $\alpha > 0$ a constant. Both these datasets are thus publicly available.

For the fertility dataset, table \ref{taba} shows minus the log-likelihood, the Akaike Information Criterion (AIC) and the values of $\alpha, r_1$ and $r_2$ for the Poisson and b-Poisson models.
The value of $r_1$ was fixed at unity, as the fit is very insensitive to the value of $r_1$. 
Note that Winkelmann's RP-$\gamma$ distribution (Winkelmann, 1995) gave $-\ell=2078.22$, so the fit here is slightly better
with $-\ell=2073.72$. The COM-Poisson fit is also slightly worse with $-\ell=2077.87$.
The COM-Poisson parameter $\hat{\nu}=1.43 \pm 0.067$, showing underdispersion.
The value of $r_1+r_2=1.63$ for the b-Poisson fit also shows that the distribution is underdispersed.

The fitted covariate values are compared with Winkelmann's results in table \ref{tabb}. They are very similar.

\begin{table}[!htb]
\caption{\label{taba}Fits of Poisson and r-Poisson models to the fertility data.}
\centering
\fbox{%
\begin{tabular}{lccccc}
Model & $-\ell$& AIC & $\alpha$ & $r_1$ & $r_2$ \\ \hline
Poisson&2186.78&4375.15&2.38 (.044) &-&- \\ 
b-Poisson&2176.81&4375.62&3.41 (.15)& 1 (0) & 0.425 (.059) \\ 
Poisson+covs&2101.80&4225.60&3.15 (.96) &-&-\\ 
b-Poisson+covs&2073.72&4171.44&5.13 (1.33) & 1 (0) & 0.630 (.051) \\ 
\end{tabular}}
\end{table}
\begin{table}
\caption{\label{tabb}Fits of RP-$\gamma$ and r-Poisson models to the fertility data with covariate regression.}
\centering
\fbox{%
\begin{tabular}{lllll} 
&RP-$\gamma$ &&r-Poisson & \\ \hline
$-\ell$ &2078.22 &-& 2073.72&-\\ \hline
Variable & Coeff. & se & Coeff. & se \\ \hline
German&-.190&      0.059&  -.20&      0.062 \\ 
Yrs schooling&0.032 & 0.026&   0.034&  0.028 \\   
Voc training&-.14&      0.036&  -.15 &      0.038 \\  
University&-.15 &     0.13&   -.16&      0.137 \\    
Catholic&0.21   &   0.058   &0.21 &      0.0611 \\   
Protestant&0.11 &     0.062& 0.11 &      0.066 \\        
Muslim&0.52   &   0.070&  0.55&      0.073 \\ 
Rural&0.055&0.031   &0.059&  0.033 \\  
Year of birth&0.0023&  0.0019&   0.0020&  0.0020 \\ 
Age at marriage&-.029&  0.0053&   -.030&  0.0056 \\     
\end{tabular}}
\end{table}

Turning to the `affairs' dataset, table \ref{tabc} shows minus the log-likelihood, \etc. 
The fit with covariates with $-\ell=696.3$ is somewhat better than for the negative binomial model, which gave $-\ell=728.1$.
The COM-Poisson also fitted worse, with $-\ell=906.4$, and with the parameter $\hat{\nu}=0.0138 \pm 0.006$, showing overdispersion.
One might anticipate that the COM-Poisson distribution would not fit well from figure \ref{figa}. The mean number of affairs is 1.456, standard deviation $\sigma=3.299$, giving a coefficient of dispersion of 7.475.
The b-Poisson can reach about 12.9 for this mean, and the COM-Poisson only 2.22.
\begin{table}
\caption{\label{tabc}Fits of Poisson and b-Poisson models to the extramarital affairs data.}
\centering
\fbox{%
\begin{tabular}{lccccc} 
Model & $-\ell$& AIC & $\alpha$ & $r_1$ & $r_2$ \\ \hline
Poisson&1709.72 & 3421.44&1.456 & 1 (0) & - \\  
b-Poisson&726.96& 1459.91&11.83 (0.96) & .0028 (.00023) & .078 (.0146) \\ 
Poisson+covs&1375.50& 2769.01& 9.13 (3.8) & 1(0) &- \\ 
b-Poisson+covs&698.30 & 1418.60 & 2333 (1442) & .00363 (.00066) & 0.207 (.0166) \\ 
\end{tabular}}
\end{table}
\begin{table}
\caption{Fitted parameter values for the b-Poisson model of the extramarital affairs data.
Standard errors are given in parentheses. The last column is the p-value for a test that the regression coefficient is zero.
\label{tabd}}
\centering
\fbox{%
\begin{tabular}{lll} 
Variable & Coeff & p\\ \hline
Gender & -.112 (.203) & .58 \\ 
Age & -.0337 (.0119) & .0044 \\
Years married & .0883 (.0281) & .0017 \\ 
Children?& -.347 (.305) & .254 \\ 
Religious?& -.324 (.0803) & .00006 \\ 
Educ. level & -.0047 (.0348) & .894 \\ 
Occupation & .0782 (.0651) &.181 \\ 
Rating & -.479 (.0854) & $<$ .0001 \\ 
\end{tabular}}
\end{table}
Only a few of the covariates are significant, and these findings agree closely with Hilbe (2011).
The aim here has been to show that the b-Poisson distribution can feasibly be used to fit data; if looking at these datasets in earnest one would for example probably use 
the bootstrap to obtain a more robust standard error on regression coefficients.
\section{Conclusions}
A way to generalize any discrete distribution has been proposed, that adds one or two extra parameters.
The stochastic process at the heart of the generalization can be self-exciting or self-damping,
resulting in overdispersion or underdispersion relative to the parent distribution.
The existence of this stochastic process gives the distributions a probabilistic basis whereby events (such as birth of a child) can facilitate or suppress the occurrence of further events.

Starting the stochastic process (which is a Markov chain with an embedded renewal process) at equilibrium leads to a simple formula for the mean in terms of the mean of the parent distribution.
This, plus the ability to compute probabilities, allows inference about the \role\  of covariates to be done in a straightforward way.
Generation of random numbers is also straightforward.

Computation of probabilities for the distributions is feasible and quite simple. A faster method has also been developed, suitable for computing individual probabilities, as required by
likelihood-based inference. Given $m$ as the highest non-tiny probability for the parent distribution, the time to compute the $i$th probability is $O(m-i)$.
This is slightly faster than the computing time needed for the COM-Poisson distribution (\eg Shmueli {\em et al}, 2005), a popular distribution for modelling over and underdispersed data.

Fits to two well-travelled datasets were good, and better than standard alternatives such as Winkelmann's RP-$\gamma$ distribution (underdispersion), the negative binomial (for overdispersion),
and the COM-Poisson distribution.
This is encouraging, but naturally one should not read too much into performance on two examples.

Further work could include an attempt to obtain a bivariate distribution of this type by considering a 3-sided game.
Code to generate random numbers,  and to compute a single probability quickly is available in the supplementary materials.
\section*{Acknowledgement}
I am grateful to Prof. Philip Scarf for introducing me to his model of basketball.

\section*{Appendix A: probabilities for various values of $N$}
These were derived using the method of symbolic computation described in the text, and checked manually.

\begin{align*}
N&=1\\
p_{  0}&=(r_1+r_2)^{-1}r_2\\
p_{  1}&=(r_1+r_2)^{-1}r_1\\
\end{align*}

\begin{align*}
N&=2\\
p_{  0}&=(r_1+r_2)^{-1}\{(1-r_1)r_2\}\\
p_{  1}&=(r_1+r_2)^{-1}\{     2r_1r_2\}\\
p_{  2}&=(r_1+r_2)^{-1}\{r_1(1-r_2)\}\\
\end{align*}

\begin{align*}
N&=3\\
p_{  0}&=(r_1+r_2)^{-1}\{(1-r_1)^{     2}r_2\}\\
p_{  1}&=(r_1+r_2)^{-1}\{     2r_1(1-r_1)r_2+r_1r_2^{     2}\}\\
p_{  2}&=(r_1+r_2)^{-1}\{r_1^{     2}r_2+     2r_1r_2(1-r_2)\}\\
p_{  3}&=(r_1+r_2)^{-1}\{r_1(1-r_2)^{     2}\}\\
\end{align*}

\begin{align*}
N&=4\\
p_{  0}&=(r_1+r_2)^{-1}\{(1-r_1)^{     3}r_2\}\\
p_{  1}&=(r_1+r_2)^{-1}\{     2r_1(1-r_1)^{     2}r_2+     2r_1(1-r_1)r_2^{     2}\}\\
p_{  2}&=(r_1+r_2)^{-1}\{r_1^{     2}(1-r_1)r_2+     2r_1^{     2}r_2^{     2}+     2r_1(1-r_1)r_2(1-r_2)+r_1r_2^{     2}(1-r_2)\}\\
p_{  3}&=(r_1+r_2)^{-1}\{     2r_1^{     2}r_2(1-r_2)+     2r_1r_2(1-r_2)^{     2}\}\\
p_{  4}&=(r_1+r_2)^{-1}\{r_1(1-r_2)^{     3}\}\\
\end{align*}

\begin{align*}
N&=5\\
p_{  0}&=(r_1+r_2)^{-1}\{(1-r_1)^{     4}r_2\}\\
p_{  1}&=(r_1+r_2)^{-1}\{     2r_1(1-r_1)^{     3}r_2+     3r_1(1-r_1)^{     2}r_2^{     2}\}\\
p_{  2}&=(r_1+r_2)^{-1}\{r_1^{     2}(1-r_1)^{     2}r_2+     4r_1^{     2}(1-r_1)r_2^{     2}+r_1^{     2}r_2^{     3}+\\
&     2r_1(1-r_1)^{     2}r_2(1-r_2)+  2r_1(1-r_1)r_2^{     2}(1-r_2)\}\\
p_{  3}&=(r_1+r_2)^{-1}\{r_1^{     3}r_2^{     2}+     2r_1^{     2}(1-r_1)r_2(1-r_2)+\\
&     4r_1^{     2}r_2^{     2}(1-r_2)+     2r_1(1-r_1)r_2(1-r_2)^{     2}+r_1r_2^{     2}(1-r_2)^{     2}\}\\
p_{  4}&=(r_1+r_2)^{-1}\{     3r_1^{     2}r_2(1-r_2)^{     2}+     2r_1r_2(1-r_2)^{     3}\}\\
p_{  5}&=(r_1+r_2)^{-1}\{r_1(1-r_2)^{     4}\}\\
\end{align*}

\end{document}